\documentclass[twocolumn,english,notitlepage,superscriptaddress,nofootinbib]{revtex4-2}
\usepackage[T1]{fontenc}
\usepackage[latin9]{inputenc}
\usepackage{float}
\usepackage{amsmath}
\usepackage{amssymb}
\usepackage{graphicx}
\usepackage{wasysym}
 \usepackage{color}
\makeatletter

\DeclareMathOperator*{\Ntot}{N^{tot}}

\newcommand{\affone}{Department of Physics and Astronomy, University College London, Gower Street, WC1E 6BT London, United Kingdom.}

\makeatother

\begin{document}
\title{Nanoparticle arrays levitated in a cavity for quantum sensing}
\author{J.H.  Iacoponi} 
\affiliation{\affone} 

\author{M. Rademacher} 
\affiliation{\affone} 

\author{T.S. Monteiro}
\affiliation{\affone}

\begin{abstract}
Levitated nanoparticles are being investigated as ultrasensitive quantum sensors of forces and accelerations,  with applications ranging from fundamental physics phenomena such as dark matter or quantum gravity to real world applications.   Attention is now turning to multiparticle regimes,  and  an important question is whether  collective effects offer advantages for sensing.
We investigate here the spectral characteristics of collective motion of  $N $ trapped nanoparticles interacting via the optical mode of a cavity.   We find the collective motion typically exhibit two generic spectral  features:   a broad spectral feature,  the collective bright mode (CBM) which has been previously studied; but we find also a new structure of sharp peaks , the mechanical mode comb (MMC).  We  can  describe all the detailed spectral features of the system,  with a simple closed-form expression,  by reducing the motion to a 1D generic collective mode which is {\em non-Hermitian}.  We  show that the MMC is more advantageous than the usual CBM for increased sensitivity in force sensing.  We find that the mechanical comb can  autonomously  repair loss of `teeth' due to particle loss,  a feature that may offer robustness in sensing.
\end{abstract}
\maketitle


\section{Introduction}
Levitated nanoparticles in high vacuum  ~\cite{millen2020optomechanics}  offer near-perfect decoupling from environmental noise and decoherence.  They offer a platform for exploring quantum behavior on the macroscopic or mesoscopic scale \cite{Stickler2020}.  Cooling to  quantum (or near) ground state of a single mode of the motion  was  achieved recently  both by active control~\cite{Magrini2021real,Novotny2021quantum}  or via the optical mode of a cavity~\cite{delic2020cooling,Marin2022,Novotny2023}.  

Levitation-based platforms offer also technologically important applications due to their exceptional sensitivity to weak forces and acceleration.  Experiments are undergoing rapid  development,  driven by  applications in fundamental physics such the search for dark matter  \cite{Kilian2024,Geraci2010,Arvanitaki2013,Carney2021a,Carney2021b} as well as real world applications such as  communication with under sea vessels  \cite{Fu2024}.  Sensing of forces as low as the zeptoNewton scale has already been demonstrated~\cite{Geraci2016,hebestreit2018sensing,dark_matter_3, liang2023yoctonewton, hempston2017force,timberlake2019static}.   

One to two particle set-ups have been investigated for quantum cooling and sensing \cite{ MTTM2020, MTTM2021, Marin2021,Pontin2023, PontinNP2023, Novotny2023,Gosling2024, Delic2022, Novotny2pcle,Pontin2025PRL}.  
Now multiparticle dynamics, of $N > 2$ particles, is of growing  interest, opening the way to sensing based on nanoparticle arrays,  analogous to atomic based quantum sensing \cite{YeZoller2024} , but now with higher masses.  

\begin{figure}[ht!]
{\includegraphics[width=2.8 in]{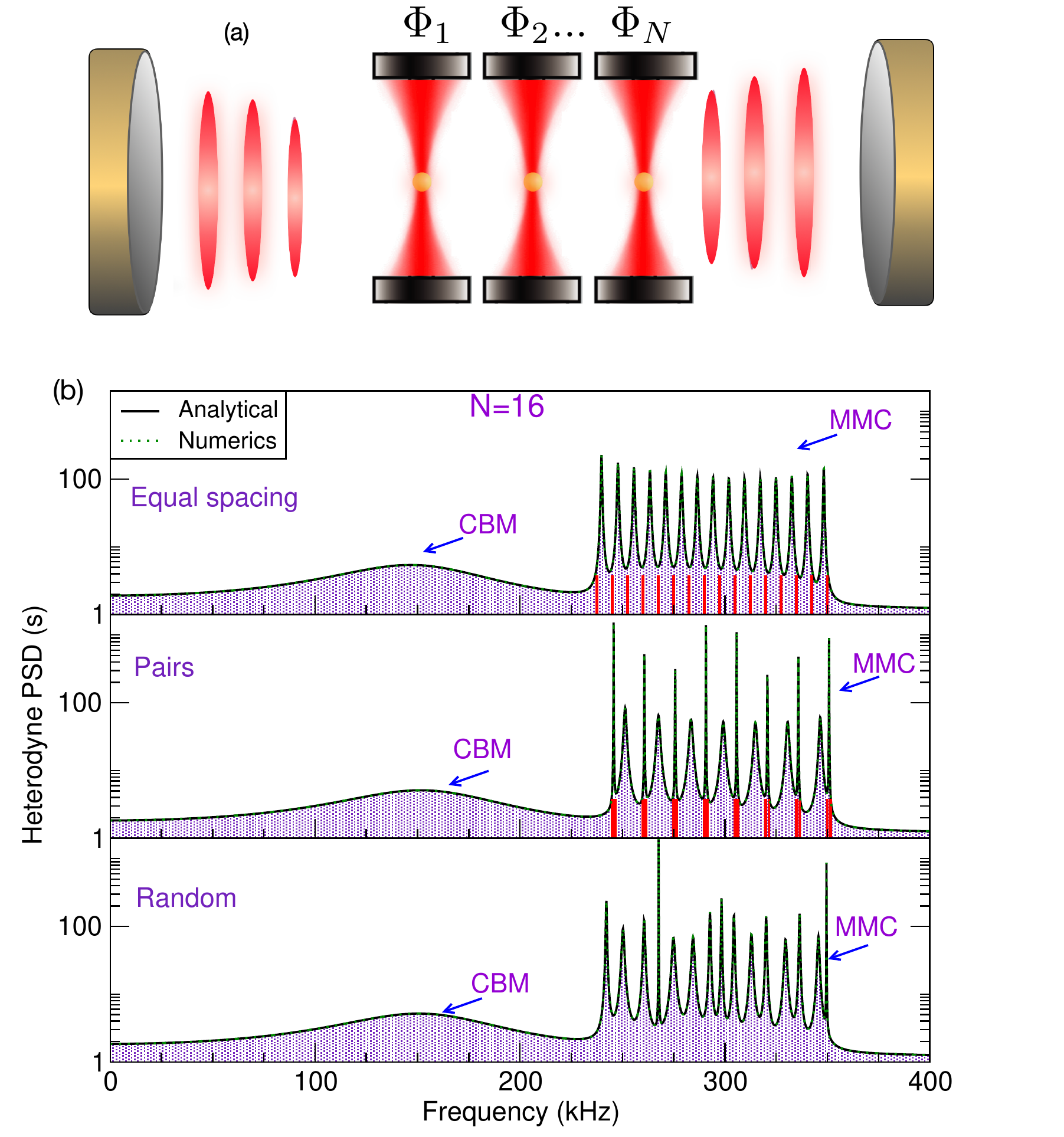}} \caption{ \textbf{(a)} An array of $N$ nanoparticles,  levitated in optical  tweezer traps,  have motional frequencies $\omega_j = \omega_1,\omega_2...\omega_N$. The array is in a cavity,  but not necessarily in a linear configuration.  The tweezer fields have phases $ \Phi_j$ \textbf{(b)} Shows the heterodyne  output.  For identical particles,  $\omega_j\equiv \omega_1$  there is only a very broad spectral feature, the collective bright mode (CBM).  In a realistic setting, there will be a spread in $\omega_j$  (red lines); we show that, in this case,  a new generic spectral feature, the mechanical mode comb (MMC),  appears.    We illustrate the MMC for the cases where the unperturbed frequencies are (i) equally spaced (ii) pairs of near degenerate modes: here, an almost equidistant pattern of alternating broad/narrow features arises (iii) randomly distributed. The spectra are perfectly described  by Eq.(\ref{Shet})(black line).}
\label{Fig1}
\end{figure}

Collective modes have been investigated in optomechanics \cite{Florian2011},  and briefly reviewed in \cite{Gieseler2018}, see references therein.
 A known phenomenon in atomic, molecular and optical physics is that for an ensemble with $j = 1, 2, \ldots, N$ oscillators that all couple with equal strength $g_j \equiv g$ to a common `bus' mode $\hat{a}$, but that otherwise have no direct interaction, there results an effective enhancement of the coupling strength to $\hat{a}$, which couples only to a center of mass mode $\hat{Q}$ with effective strength $g\sqrt{N}$.  Levitation-based force sensing for an array of identical nanoparticles (degenerate modes) has been analysed \cite{YaoLi2023} to consider this collective enhancement in coupling.
 
\begin{figure}[ht]
{\includegraphics[width=3.  in]{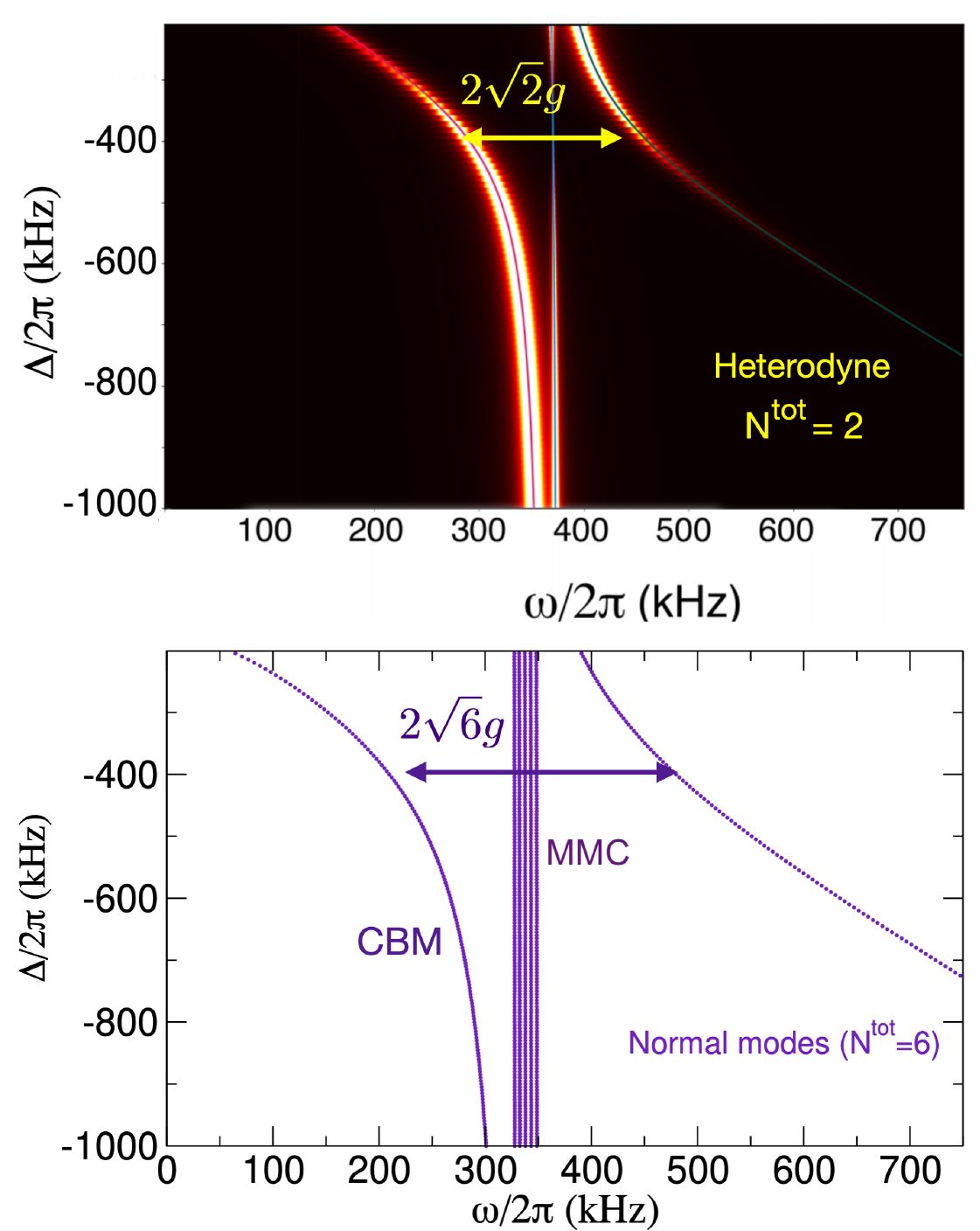}} \caption{ Shows that the CBM/MMC features,  in systems with $\Ntot\geq  2 $  mechanical modes,  may be viewed as the multi-particle analogue of the bright-dark mode state spectra  analysed in \cite{MTTM2021} and investigated in recent experiments \cite{Marin2021,Novotny2023}.  Compares the behavior between $\Ntot= 2 $ (upper panel) and $\Ntot=6$ (lower panel) 
for the case where $g_j \equiv g, \Phi_j=0$ and thus the collective mode is not only real,  but corresponds to the center of mass coordinate. The CBM is the lower frequency branch of  an avoided crossing of width $2\sqrt{\Ntot} g$.  The MMC in turn is  a feature that results from  lifting the degeneracy of dark modes,  enabling some mixing with the bright collective mode.  }
\label{Fig2}
\end{figure}

But while an atomic array might be modelled with $N$ identical ($\omega_j \equiv \omega_m, g_j\equiv g$) oscillators, in practice no two experimentally trapped levitated nanoparticles are identical.
Even within the same prepared sample of nanoparticles, fluctuations in their radius lead to different resonant frequencies and couplings to the  cavity optical mode.
Furthermore, the particles are in distinct tweezer traps, each with arbitrarily different phases $\Phi_j$.
Thus we consider a \textit{realistic} array of nanoparticles, with an arbitrary distribution in $\omega_j, g_j$ and $\Phi_j$ -- the unperturbed resonant frequencies, couplings and phases respectively -- and with an arbitrary number of modes.

 Our first key finding is that for the non-ideal array, we have not only the broad spectral feature from the previously studied collective bright mode (CBM), but there is also a new and generic spectral feature comprising a `comb' of peaks that are individually narrow,  but span a continuous frequency range.
We term this feature the mechanical mode comb (MMC) and investigate its properties and potential for force sensing.
In particular we conclude that the very strong damping and coupling associated with the CBM is not as advantageous for sensing as the MMC.
Specifically, for optomechanical sensing of external forces, the key sources of error are back-action and imprecision noises.
It is well established that a strongly peaked force susceptibility plays an important role in suppressing imprecision noise errors near resonant mechanical frequencies, and here the effective susceptibility near the MMC is advantageous.

Our second key finding is that, even in the most general case, there is a quasi-1D analysis that yields cavity output spectra in agreement with numerical spectra.
This greatly facilitates the signal-to-noise analysis of experimental spectra, as we can always write an analytical expression for the measured optical signal for arbitrary array configurations and experimental parameters.
This quasi-1D analytical form is possible because, even in the most general case, the optical field couples to a single collective mode of the form:

\begin{align}
\label{nonHerm}
     \hat{Q}(\omega)
     \equiv
     \frac{1}{\bar{g}}
     \sum_j
         g_j
         e^{-i\Phi_j}
         \hat{q}_j(\omega)
\end{align}
with $\bar{g} \equiv \sqrt{\sum_j g^2_j}$.
We see that  $\hat{Q} \neq \hat{Q}^\dagger$ and  the mechanical displacements are -- in the general case where the tweezer traps have arbitrarily different phases -- complex and non-Hermitian.
For analysis of force sensing, the physical meaning of the non-Hermitian collective motion operator $\hat{Q}$ is not essential,  as we are simply considering $\hat{Q}$ as an effective transducer that converts external force fluctuations into fluctuations of the measured output signal.
When $\Phi_j = 0$ and $g_j = g$ for all $j$,  it reduces to the well-known collective center of mass mode $\hat{Q} \equiv \sum_j \hat{q}_j/\sqrt{N}$ of the array.
For $\Phi_j = 0$ and unequal $g_j$, the mode is real and represents simply a more general form of the center of mass mode.

In Section II, we begin by presenting a (numerical) study of the key spectral features, the CBM and MMC.
The MMC is completely unrelated to optical combs well studied in laser physics.
However, we show that it is the $N$-particle generalisation of the bright/ dark mode phenomena investigated theoretically and experimentally for a one-particle, two-mode optomechanical system.
There are other ways in which these apparent imperfections in the nanoparticle ensemble (in the sense of non-identical nanoparticles) might even be turned into an advantage for sensing, and we identify numerically a robustness to loss of particles in the array: the resultant gap in the comb `self-repairs' as neighbouring `teeth' move and widen to fill it.
 
In Section III we present our quasi-1D model.
We show that it yields near exact agreement with the full numerics for the optical spectrum.

In Section IV we present our analysis of the force sensing using the 1D model.
We obtain the form of the back-action and imprecision noises, as well as the effective force susceptibility. 

In Section V we discuss our results and conclude.

\section{Spectra of  levitated arrays}
In Fig.(\ref{Fig1}a) we illustrate an array of $N$ nanoparticles that are held in individual optical tweezer traps within an optical cavity.
While the tweezer traps parameters and nanoparticle properties determine the mechanical frequencies $\omega_j$, the optical cavity mode provides a range-independent effective coupling between the nanoparticles.
Thus there is no requirement for a spatially linear configuration of the nanoparticles. 
In recent quantum cooling experiments \cite{delic2020cooling,Marin2022,Novotny2023} the cavity was undriven, leaving the optical mode to be populated exclusively by coherent scattering from the nanoparticles.

It is important to note that number of dynamical modes of interest is $\Ntot \neq N$; that is, in general, the number of dynamical modes can exceed $N$.
Considering 3D center of mass motion (the $x,y,z$  degrees of freedom),  in principle,  $\Ntot = 3N$ where $N$ is the number of trapped particles; in 2D,  $\Ntot = 2N$.  More recent experiments \cite{Novotny2pcle} on two particles trapped in two tweezers found only cavity-mediated coupling was significant.
For tweezers in phase, intra- and inter-particle couplings take exactly the same form,  so whether we consider $\Ntot = 3N$, $2N$ or $N$ does not affect our results.
For the illustrations, we mainly focus on parameters relevant to recent experimental setups that that allowed quantum cooling in one and two modes \cite{delic2020cooling, Marin2022, Novotny2023} of a single particle.

For non-spherical particles, rotational and librational modes are also of great interest, and most of our results are generic, but here we focus primarily on the quantum center of mass coordinate $\hat{q}_j$, as for typical experiments other types of modes are well separated in frequency.
For example, in a typical coherent scattering cavity setup, motion in the $\hat{x}, \hat{y}$ directions -- transverse to the tweezer's propagation axis -- is typically at much higher frequency and is more strongly coupled to the light, than motion in the propagation axis $\hat{z}$.
Without loss of generality, for convenience we mainly choose $\Ntot = 2N$ or $\Ntot = N$ in our numerical examples (which use typical experimental parameters) as we have investigated 1D and 2D mode dynamics and force sensing theoretically and experimentally in previous studies \cite{MTTM2020, MTTM2021, Pontin2023, Gosling2024}.

\subsection{Multiparticle and multimode spectra}

\subsubsection{ Hamiltonian and equations of motion:}

We model the array by means of a linearised Hamiltonian of the form:

\begin{align}
\label{Hlin}
    \hat{H} / \hbar
    =
    & -\Delta \hat{a}^\dagger \hat{a}
    +
    \sum_j
    \omega_j \hat{b}_j^\dagger \hat{b}_j
    \nonumber
    \\
    & -
    \sum_j
    g_j
    (
        \hat{a} e^{i\Phi_j}
        +
        \hat{a}^\dagger e^{-i\Phi_j}
    )
    (
        \hat{b}_j
        +
        \hat{b}_j^\dagger
    )
\end{align}
where $\hat{a}$ ($\hat{a}^\dagger$) is the annihilation
(creation) operator for the optical mode and $\hat{b}_j$ ($\hat{b}_j^\dagger$)
for mechanical mode $j$.
The corresponding mechanical mode displacement operators are $\hat{q}_j = \hat{b}_j^\dagger + \hat{b}_j$.
$\Delta$ is the detuning in frequency between the input laser and the cavity mode, while $\omega_j$ and $g_j$ are the natural frequency and optical mode coupling strength, respectively, of the $j^\text{th}$ mechanical
mode.

\begin{figure}[ht!]
{\includegraphics[width=3.3in]{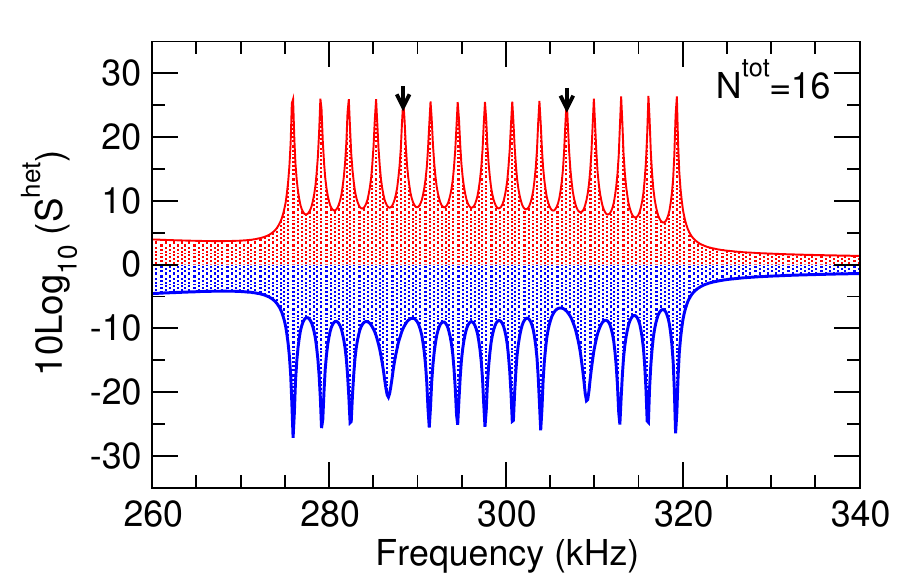}} \caption{  Robustness of the mechanical mode comb (MMC) to particle loss (loss of its `teeth').
    Upper comb shows the heterodyne spectrum from a 16-particle setup with regularly spaced mode frequencies (with uniform coupling $g/(2\pi) =30$ kHz, and frequency spread $\Delta\omega / (2\pi)= 45$kHz).
    Lower comb shows the spectrum (reflected about horizontal axis) if the two nanoparticles corresponding to the black arrows are knocked out of the array: here, the neighbouring teeth of the comb displace and broaden so as to largely repair the gap; whereas an independent particle array might leave a near zero-signal gap at the missing frequencies.}
\label{Fig3}
\end{figure}

In frequency space, and including dissipative terms, the Hamiltonian gives rise to a set of $\Ntot+2$ quantum  equations of motion
for the optical  modes:
\begin{eqnarray}
\hat{a}(\omega) &=& \ \  \ \chi_c (\omega)  \left[\textrm i   \sum_j (g_j e^{-i\Phi_j}\hat{q}_j(\omega) )
+ \sqrt{\kappa} \tilde{a}_{{in}} \right]\nonumber \\
\hat{a}^\dagger(\omega) &=&  \chi^*_c (-\omega) \left[ - \textrm i   \sum_j (g_j e^{i\Phi_j}\hat{q}_j(\omega) )
+ \sqrt{\kappa} \tilde{a}^\dagger_{{in}} \right]
\label{aopt1}
\end{eqnarray}

and for the mechanical displacement modes:
\begin{eqnarray}
\hat{q}_j(\omega) &=& \sum_j \sqrt{\Gamma}_j  [\chi_j(\omega) \tilde{b}_{j,in}+ \chi^*_j(-\omega)\tilde{b}^\dagger_{j,in}] \nonumber \\
&+& \textrm ig_j  \eta_{j}(\omega)[\hat{a}(\omega)e^{i\Phi_j}+\hat{a}^\dagger(\omega) e^{-i\Phi_j}].
\label{qeq}
\end{eqnarray}

The first terms  contain  $ \tilde{b}_{j,in},\tilde{b}^\dagger_{j,in} $ ,   thermal noise operators (details in appendix) corresponding to the $T_B=300$K thermal Gaussian bath due to Brownian motion of background gas molecules.  They represent the response of the nanoparticle motion to the environmental thermal forces, with damping rates $\Gamma_j$.

The second set of terms  contain $\tilde{a}_{{in}}$  optical noises,  which (ideally though typically not in practice) correspond to the quantum shot noise limit.  They are associated with the optical damping rate $\kappa$.

We have also introduced the optical (cavity mode)  susceptibility $\chi_c(\omega) =\left[-i(\omega+\Delta)+\frac{\kappa}{2}\right]^{-1}$  and mechanical susceptibilities  $\chi_j(\omega)=\left[-i(\omega-\omega_{j})+\frac{\Gamma_j}{2}\right]^{-1}$.
For compactness, below we use $\eta_j (\omega)=\chi_j(\omega) -\chi^*_j(-\omega)$ and $\eta_c (\omega)=\chi_c(\omega) -\chi^*_c(-\omega)$.

The $\Ntot+2$ quantum  Langevin equations Eqs. (\ref{aopt1}) and ( \ref{qeq}) corresponding to $\hat{a}, \hat{a}^\dagger,  \hat{q}_1, \hat{q}_2 ...\hat{q}_N $ may be solved numerically in frequency space.

We may readily use the solution for $\hat{a}(\omega)$ to obtain the cavity output field from input-output theory:
\begin{equation}
\hat{a}_{out}(\omega) =  \tilde{a}_{{in}} - \sqrt{\kappa} \hat{a}(\omega) 
\label{inout}
\end{equation}

 to obtain the power spectral density (PSD)  $S_{ \hat{a}_{out}\hat{a}_{out}}= \langle | \hat{a}_{out}(\omega) |^2 \rangle$  that would  correspond  to a heterodyne experimental measurement of the optical  output of the cavity.   This PSD imay be used,  with the appropriate analysis,  to infer experimentally the temperature of the nanoparticle; to ascertain if  it is in quantum regimes; and finally,  to infer the presence and magnitude of an external force.
 
 In order to calculate the optical PSD spectra,  we first solve the equations of motion by numerical diagonalization.  We can then present and identify generic spectral features of the array.  The numerics  also provide a check on the theory model that is presented later. 
 
 \subsection{Spectra of the N-particle array: Numerics}
 
 The numerics allow us already to  identify  two classes of spectral structures that are quite generic,  so we can illustrate them  by considering the simplest case of an ensemble for which $\Phi_j=0$ and $g_j\simeq g$.  

The $\Ntot$ modes span a frequency range $\Delta \omega$ about a mean, so  $\omega_j \in [\omega_m \pm \Delta \omega/2]$,   and the frequencies have mean spacing $\delta= \Delta \omega/\Ntot$.
There is no requirement for precisely equal couplings,  just that the couplings $g_j \sim g$  are of similar order. 
We investigate the regime   $g \sim \Delta \omega$.

\subsubsection{ Collective Bright Mode and Mechanical Mode Comb}

In Fig.(\ref{Fig1}b)  we show that the cavity output exhibits the very broad spectral feature (collective bright mode or CBM) that 
 is also seen in the degenerate case \cite{YaoLi2023}.  But now we introduce and investigate a new feature unique to these not-perfectly-identical  levitated nanoparticle arrays,  the mechanical mode comb (MMC),   that is the central focus of this work.  
 
The comb shows orders of magnitude variability in the peak widths.  However,  calculation of the variances of $ \hat{q}_j(\omega)$ for individual modes ( possible with the full numerics),  shows that the modes are mostly  thermalized,  with phonon occupancies $n_j =k_BT_j/(\hbar\omega_j) \sim n$ that are of similar order.  Because of mode hybridisation,  there is not a one to one mapping between each spectral peak to a single mode.  The comb also shows the interesting feature that peaks move to mostly fill `gaps' in the teeth of the unperturbed comb.  This is clearest for pairs (middle panel): one sees that the peaks broaden and are displaced so as to fill the spectral gap.  In Fig.\ref{Fig1}(b), 
$g/(2\pi)=40$ kHz while  $ \Delta\omega/(2\pi)=100$kHz and the (red) detuning $-\Delta=1.5 \omega_m$.  We consider the sideband resolved regime and take $\omega_m  \simeq 1.5 \kappa$.

In Fig.(\ref{Fig1}b) we presented the numerical spectra obtained numerically;  but also superposed the  results obtained from an effective single mode analytical model presented later in this work. The latter  gives indistinguishable results from the full numerics even for large $N$,  thus offers a versatile and insightful tool for experimental analysis of  arrays with large numbers of nanoparticles.

\subsubsection{ The CBM/MMC as bright/dark  modes. }

  In Fig.\ref{Fig2} we show that the CBM is the multiple-mode generalisation of the bright mode that was analysed in \cite{MTTM2021} then experimentally investigated in \cite{Marin2021} for $N=1$.  In  strong-coupling regimes the CBM then  appears,  above resonance ($-\Delta \gtrsim \omega_j$), as the low-frequency branch of a wide avoided crossing of width $2g\sqrt{\Ntot}$.

  One  notable feature,   that was noted above   is investigated  in calculations presented in  Fig.\ref{Fig3}: the  remarkable tendency of the MC to autonomously repair the loss of some of its teeth.  Particle loss is a perennial problem in these set-ups :  the problem of efficient experimental reloading of a new nanoparticle after a particle is lost from the optical trap,  remains to be fully addressed.  In this respect, the mechanical mode comb,  in the $g\sim \Delta\omega$ regime,  may add some robustness to particle loss.

\section{ Spectra of the N-particle array: 1D  model }

For the general,  many particle (or many mode) case,  we can readily re-write the equation for the optical mode in frequency space,  Eq.(\ref{aopt1}),  in the form:
\begin{equation}
\hat{a}(\omega) =  \chi_c (\omega)  [\textrm i \bar{g}  \hat{Q}(\omega) + \sqrt{\kappa} \tilde{a}_{{in}}]\
\label{aQ}
\end{equation}
where we introduce the general form of the collective mode,  for arbitrary $\Phi_j,\omega_j,g_j, \Gamma_j$: 
\begin{equation}
 \hat{Q}(\omega)  \equiv  \sum^{j=N^{tot}}_{j=1}  \frac{g_j}{\bar{g}} e^{-i\Phi_j}  \hat{q}_j .
\end{equation}
The summation is over all $N_{tot}$ motional modes.  For  1D set-ups,  it would simply  be over all particles, as $N=N^{tot}$. For simplicity,  below we use $N$  in our expressions,  but for higher dimensional set-ups,  the summations should be over $N_{tot}$.

We can then sum over Eqs. (\ref{qeq}) to re-write the collective mode in terms of the optical mode and thermal  noises:

\begin{eqnarray}
\hat{Q}(\omega) &= & \frac{1 }{\bar{g}}  \sum_j  \sqrt{\Gamma_j}  g_j e^{i\Phi_j} \left[ \chi_j(\omega) \tilde{b}_{j,in}+ \chi^*_j(-\omega)\tilde{b}^\dagger_{j,in}\right] \nonumber\\
                              & + &  \frac{i }{\bar{g}}  \sum_j   g^2_j \eta_j (\omega) (\hat{a}^{\dagger}e^{-i\Phi_j}+ \hat{a} e^{i\Phi_j}) 
\label{Qform}
\end{eqnarray}

We note that $\hat{Q}(\omega) \neq \hat{Q}^\dagger (\omega) $ thus this operator is not Hermitian.
It may however be expressed in terms of the optical and thermal noises driving the dynamics.

Thus we re-arrange Eqs.(\ref{aQ}) and (\ref{Qform}) to express the collective mode in terms of the environmental noise forces driving it :

\begin{eqnarray}
  \hat{Q}(\omega)&=& \sum_j \frac{\sqrt{\Gamma}_j  }{ \mathcal{M}_{N}} \mathcal{F}_{j}(\omega)\left[ \chi_j(\omega)   \tilde{b}_{j,in} +  \chi_j^*(-\omega)   \tilde{b}^\dagger_{j,in} \right] \nonumber \\ 
  &+& \frac{ \sqrt{\kappa} } { \mathcal{M}_{N} }  i \ \left[ \mathcal{F}_{a}(\omega)  \chi_c (\omega)  \tilde{a}_{{in}} +\mathcal{F}_{a^\dagger}(\omega)  \chi_c^* (-\omega) \tilde{a}^\dagger_{{in}} \right]  \nonumber \\
\label{Qfnoises}
\end{eqnarray}

The prefactors $ \mathcal{F}_{a},\mathcal{F}_{a^\dagger},\mathcal{F}_{j}$ are scalar functions of $\omega$ which denote the array corrections:

\begin{eqnarray}
\mathcal{F}_{j}(\omega) &=&  \frac{g_j}{\bar{g} } e^{-i \Phi_j}  \left[1 +\chi_c^*(-\omega) \mathcal{H}_j(\omega) \right]\nonumber \\
\mathcal{F}_{a}(\omega) &=& \frac{1}{\bar{g}}  \sum_{j=1}^{j=N} g^2_j \eta_j (\omega) (1+ \chi_c^*(-\omega) \mathcal{H}_j(\omega))\nonumber\\
\mathcal{F}_{a^\dagger}(\omega) &=& \frac{1}{\bar{g}}  \sum_{j=1}^{j=N} g^2_j \eta_j (\omega)  e^{-2i \Phi_j}  
\label{Fa}
\end{eqnarray}
while  the  normalization function,
\begin{equation}
 \mathcal{M}_{N}(\omega)= 1+ \sum_{j=1}^{j=N} g^2_j \eta_j (\omega) [\eta_c(\omega) +\chi_c(\omega) \chi_c^*(-\omega)\mathcal{H}_j(\omega))].
\label{MN}
\end{equation}

The prefactor
\begin{equation}  
    \mathcal{H}_j(\omega) = \sum_{k=1}^{k=N} g^2_k \eta_k(\omega) [e^{2i(\Phi_j-\Phi_k)} -1 ]
  \end{equation}

  is a function that vanishes
if $\Phi_j=0$ for all $j$, thus if all tweezers are in phase (and obviously if $j=1$).

if there is only a single particle (mode),  $j=1$,  and $N=1$,  thus ${\bar{g}}\equiv g$,  then $\mathcal{F}_{j=1}=1$ while $ \mathcal{F}_{a}=\mathcal{F}_{a^\dagger}=g \eta_{j=1}(\omega)$,  $ \mathcal{M}_{1}= 1+g^2 \eta_c(\omega)\eta_{j=1}(\omega)$,  
and the  expression reduces to the  well-studied single particle case,  a textbook example in  quantum optomechanics.

\subsection{Power Spectral Densities (PSDs)}

In this section we show how PSDs for an array may be obtained analytically for arbitrary $N$ and arbitrary 
experimental parameters from the effective 1D dynamics.
It is useful to  review briefly the  textbook example in quantum optomechanics of the single-particle coupled to a single cavity mode,  thus where  $N=1$,  $g_j\equiv g$ and $\Phi_j=0$.

\begin{figure}[ht!]
{\includegraphics[width=3.in]{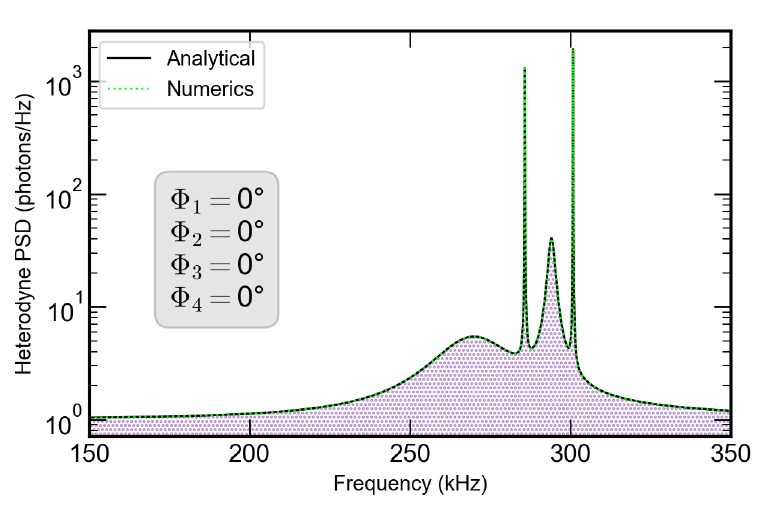}}
{\includegraphics[width=3.in]{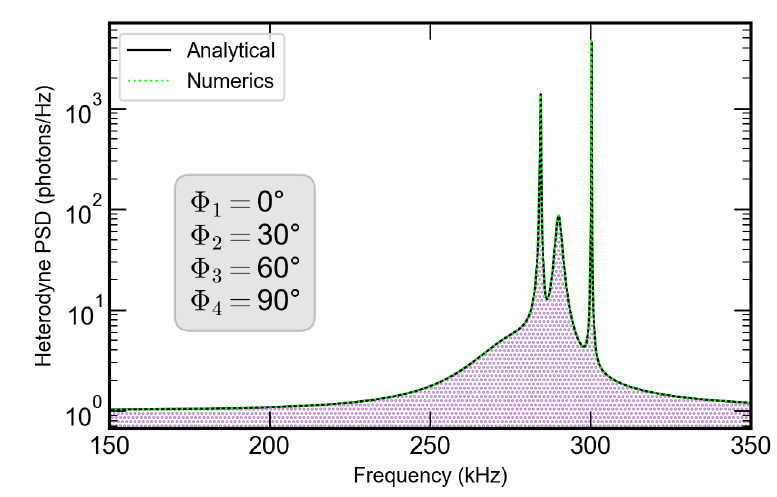}}
 \caption{ Shows comparisons between the usual numerics and Eq.(\ref{Shet}) for a test with $N=4$ particles (or modes). 
 The upper panel shows the case where all the tweezers are in phase.  In this case the mode that couples to the light is the usual -physical- centre of mass mode.  The lower panel illustrates the case where there are 4 tweezers all with different phases. In this case the displacement coordinate that couples to the light is non-Hermitian.  In both cases the results obtained from Eq.(\ref{Shet}) and numerics are indistinguishable so the much simpler Eq.(\ref{Shet}) represents a useful tool for analysis and optimisation of the effective force susceptibility in  multiparticle and multimode regimes. }
\label{Fig4}
\end{figure}

\subsubsection{PSD for a  single mode}
We re-arrange Eqs.(\ref{aopt1}) and (\ref{qeq}) to readily
 obtain the displacement in terms of the environmental  noise forces acting on it:
\begin{eqnarray}
 \hat{q}_1 (\omega) &=& \frac{\sqrt{\Gamma}  }{ \mathcal{M}_{1} (\omega)} [\chi_1(\omega) \tilde{b}_{1,in}+ \chi^*_1(-\omega)\tilde{b}^\dagger_{1,in}] \nonumber \\
&+ & \frac{ \sqrt{\kappa}}{ \mathcal{M}_{1}(\omega) } ig  \eta_1(\omega) \ [\chi_c(\omega)\tilde{a}_{in}+ \chi_c^*(-\omega)\tilde{a}_{in}^\dagger ]
\label{aoptmech}
\end{eqnarray}
noting we assume that $j=1$.  The correlators of the noise operators are known:
\begin{eqnarray}
 \langle \tilde{Y}(\omega)  \tilde{Y}^\dagger_{in}(\omega') \rangle &= &(n_Y +1) \delta(\omega+\omega ')  \nonumber \\
  \langle \tilde{Y}^\dagger(\omega)  \tilde{Y}_{in}(\omega') \rangle &= & n_Y \delta(\omega+\omega ') 
\label{corr}
\end{eqnarray}
We denote $n_{Y}=n_a$ for  the photon number of the cavity mode while $n_Y \equiv n_j$ is the phonon occupancy of the thermal environment.  In the ideal case,  the optical mode has zero temperature occupancy, thus  $n_a=0$.

For generality,  we  can take the corresponding mechanical damping to be different between nanoparticles:  $\Gamma_j $, is  related to gas pressure $P$ as $\Gamma_j \sim 15PR_j^2/(M_j{\bar{v}})$  with $R_j,M_j$ the nanoparticle radius, mass and ${\bar{v}}$ the mean gas velocity.  The Brownian bath has phonon occupancies of $n_{j}= kT_B/(\hbar \omega_j)$.   For many realistic scenarios,  with spherical nanoparticles of comparable radius,  we can for simplicity assume $\Gamma_j \equiv \Gamma$ if required.

Using the correlators from Eq.(\ref{corr}) and taking $n_a=0$, we obtain the well-known form of the optomechanical PSD for the mechanical motion $S_{q_1 q_1} (\omega)= \langle |\hat{q}_1 (\omega)|^2\rangle$:

\begin{eqnarray}
S_{q_1q_1} (\omega) &= & \frac{\Gamma_1 }{ |\mathcal{M}_{1} |^2} \left[ |\chi_{1}(\omega)|^2 (n_1+1) + |\chi_{1}^*(-\omega)|^2 n_1 \right] \nonumber \\
&+ & \frac{\kappa }{ |\mathcal{M}_{1} |^2} g^2 |\eta_{1}(\omega)|^2  |\chi_c(\omega)|^2  
\label{PSD1} 
\end{eqnarray}
with the normalization term, $ \mathcal{M}_{1}= 1+g^2 \eta_c(\omega)\eta_{1}(\omega)$.  The above expression is used widely  to model  optomechanical cooling in cavities
 (for levitated and non-levitated set-ups)  as the PSD,  integrated over frequency space,  may be used to infer the temperature of the motional mode.

\begin{figure*}[ht!]
{\includegraphics[width=5.  in]{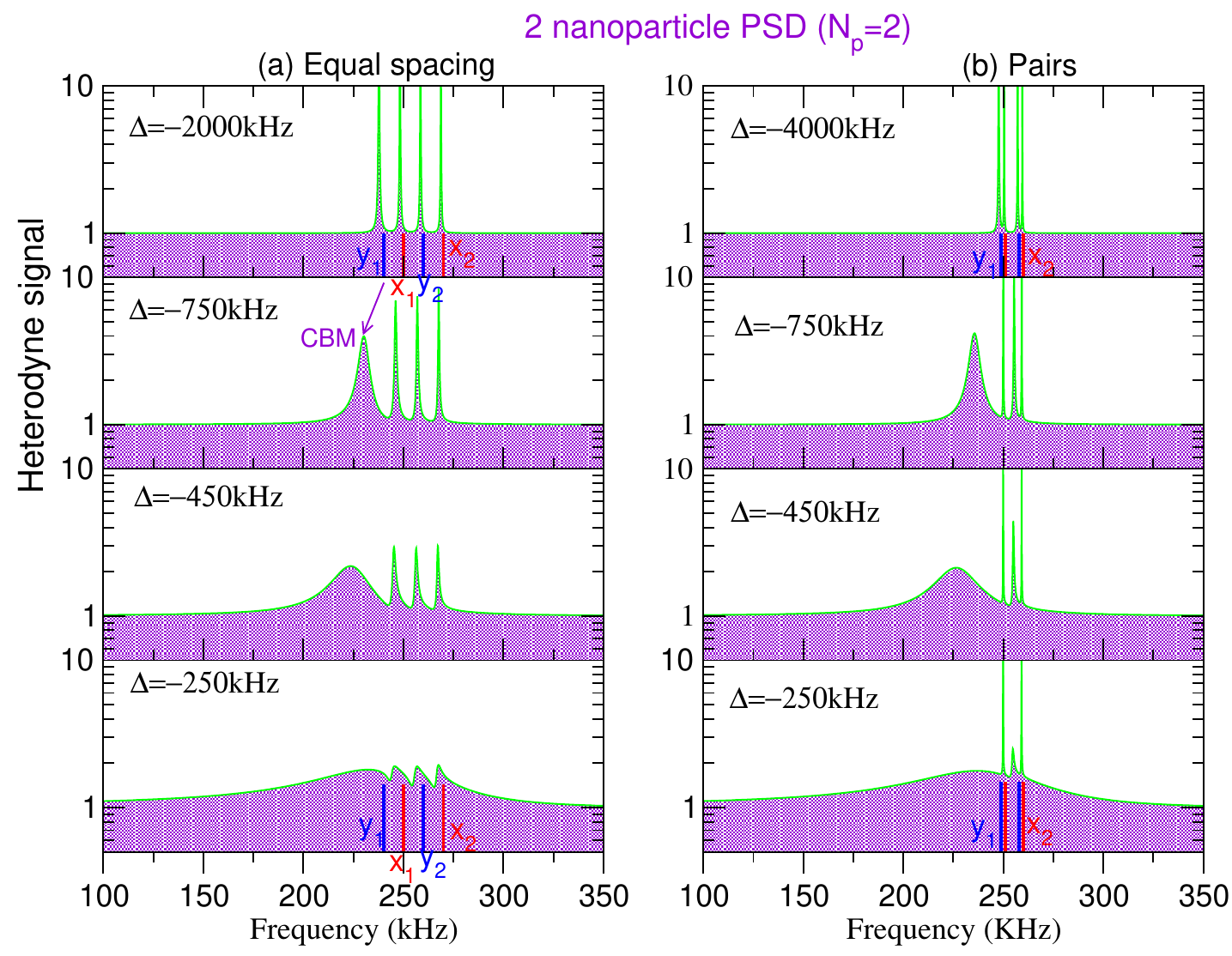}} \caption{ {\bf (a)}  Illustrates that phenomena such as collective bright mode (CBM) and  MMC formation are already accessible with current $N=2$,  2D experiments in levitated cavity optomechanics such as \cite{Novotny2pcle}.  If the detuning is far off-resonance,   ($\Delta\to \infty $),  4 narrow peaks at the tweezer frequencies are seen.  As $-\Delta$ is  tuned nearer resonance ($-\Delta \sim \omega_j$) , the lower frequency peak peels off to form the CBM.  The MMC remains relatively weakly coupled but also experiences some cooling; for sufficient  $\delta$ ( spacing between unperturbed frequencies) the MMC can even reach quantum regimes.  {\bf (b)} The tendency of the MMC to fill frequency gaps is also shown using paired unperturbed frequencies.}
\label{Fig5}
\end{figure*}

\subsubsection{PSDs for multimode arrays }

 If we compare Eq.(\ref{Qfnoises}) and its single-particle equivalent, Eq.(\ref{aoptmech}) we see we can immediately write down the array PSD without further calculation:
\begin{eqnarray}
&S_{QQ}(\omega)&= \nonumber\\
&\sum_j &  \frac{\Gamma_j }{ |\mathcal{M}_{N} |^2} |\mathcal{F}_{j}(\omega)|^2 \left[ |\chi_j(\omega)|^2 (n_j+1) + |\chi_j^*(-\omega)|^2 n_j \right] \nonumber \\
&+ & \frac{\kappa }{ |\mathcal{M}_{N} |^2}  |  \mathcal{F}_{a} (\omega)|^2  |\chi_c(\omega)|^2  
\label{PSDN} 
\end{eqnarray}

Comparison between Eq.(\ref{PSD1}) and (\ref{PSDN}) shows the remarkably similarity.  The behaviour arising from arbitrary phases $\Phi_j$ from different tweezers is subsumed in the $\mathcal{F}$ coefficients:  calculating the PSDs for arbitrary arrays reduces to simply  inserting the array correction terms
 and the normalization term $\mathcal{M}_{N}$ into the usual expressions.  A similar simplification is found for
 the PSD of the cavity output, as seen below.  This is  one of the central results of this work.

\subsection{Measured displacement spectra for an array}

Of course neither the single or N-particle displacement spectra in  Eqs.(\ref{PSD1}) and (\ref{PSDN}) are directly observed. 
They are inferred via the cavity output spectra.  Once again,  expressing the operators in terms of the noises such as in Eq.(\ref{aoptmech}) or Eq. (\ref{Qfnoises})  enables one to calculate all PSDs of interest. 
Substituting Eq.(\ref{Qfnoises}) into Eq.(\ref{aQ}) and combining with the input-output expression Eq.(\ref{inout}) we can obtain
$\hat{a}_{out} $ but for an $N$-mode array, in terms of noises and thus the corresponding PSD:

\begin{equation}
\label{Shet}
\begin{aligned}
S_{\hat{a}_{\mathrm{out}}\hat{a}_{\mathrm{out}}}
&= \frac{\kappa \bar{g}^{2} |\chi_c(\omega)|^{2}}
         {|\mathcal{M}_{N}|^{2}}
   \biggl\lbrace
   \sum_j \Gamma_j |\mathcal{F}_j|^{2}
\\
&\quad {}\times
   \Bigl[
      |\chi_j(\omega)|^{2}(n_j+1)
      + |\chi_j^{*}(-\omega)|^{2}n_j
   \Bigr]
\\
&\quad {}+
   \kappa |\chi_c(\omega)|^{2}
   |\mathcal{F}_a|^{2}
   \biggr\rbrace
\\
&\quad {}+
   |1-\kappa\chi_c(\omega)|^{2}
\\
&\quad {}+
   2\operatorname{Re}\bigl[
      \frac{\kappa\chi_c(\omega)\mathcal{F}_a}
           {\mathcal{M}_{N}}
      \Bigl(1-\kappa\chi_c^{*}(\omega)\Bigr)
   \big].
\end{aligned}
\end{equation}

 In an experiment,  the  corresponding measured PSD of the cavity output may then be used to estimate the displacement spectrum:
 
\begin{equation}
 S^{\text{meas}}_{QQ}(\omega) \approx \frac{S_{ \hat{a}_{out} \hat{a}_{out}} (\omega)}{ \kappa\bar{g}^2 |\chi_c(\omega)|^2}.
\end{equation}
by dividing by the mean coupling and optical susceptibility.

We can similarly rescale the theoretical expression; then by inspection we see that the 4 terms in each line of Eq.(\ref{Shet}) correspond respectively to 4 distinct physical contributions:
\begin{equation}
 S^{\text{meas}}_{QQ}(\omega) \equiv \sum_j \Gamma_j S_{QQ}^{j,therm} + S_{BA} + S_{\text{imp}} + S_{\text{interf}}
\end{equation}
which all map onto 1-particle terms,  but have a different form for the array.

{\em The first term, $S_{QQ}^{j, therm}(\omega)$}  denotes the effect on $\hat{Q}$ of  the thermal forces acting on the particle.  This is a  very important term as it indicates the effective force susceptibility, and thus the response if we add an external force.  
We may rewrite:
\begin{eqnarray}
  \sum_j \Gamma_j S_{QQ}^{therm,j} & \equiv &  \sum_j \Gamma_j   \frac{ |\mathcal{F}_{j}(\omega)|^2  |\chi_j(\omega)|^2 }{|\mathcal{M}_{N} |^2(\omega)}  S_{FF}^{(j, therm)} \nonumber \\
& = &  \sum_j \Gamma_j   | \chi_j^{\text{eff}}(\omega)|^2  S_{FF}^{(j, therm)}
\end{eqnarray}
where $S_{FF}^{(j, therm)}$ the thermal force spectrum on particle $j$,  and where:  
\begin{equation}
  \chi_j^{\text{eff}}(\omega)  \equiv  \frac{ \mathcal{F}_{j}(\omega)  \chi_j(\omega) }{ \mathcal{M}_{N} (\omega)}
  \end{equation}
   is the effective force susceptibility.  This is a key result: the extraordinary spectral complexity of the array and multimode set-up is 
   completely contained in the easy to evaluate scalar prefactor $ \mathcal{F}_{j}(\omega)$ and modified normalization 
   $\mathcal{M}_{N} (\omega)$.

In general the environmental noises are broad band or white noise; and operating at high vacuum we may take $\Gamma \to 0$ to eliminate the effect of gas collisions.  But if we consider a  deterministic external force $F^{ext} (\omega)=\sum_j  F_j^{ext}(\omega)$ 
 the above susceptibility allows analysis of  its signal in the optical spectrum. \\

{\em The second and third terms},  $ S_{BA} +S_{\text{imp}}$ are the back action and imprecision noises respectively.
They represent the key sources of error in optomechanical sensing.  Minimising  the sum leads,  in ideal conditions to the so-called Standard Quantum Limit. 

It is important to note that while in Eq.(\ref{Shet}) the form of  $S_{\text{imp}}(\omega) $ assumes the ideal quantum shot noise,  in general,  there are other contributions to $S_{\text{imp}}$, arising from post-processing,  electronics as well as laser phase noises so the imprecision  contribution may be considerably larger. \\

{\em The fourth term} $S_{\text{interf}}$ which is the interference between the input noise and the optical back action terms is included in all calculations but not discussed extensively.  This term is of wide interest in optomechanics as it leads to ponderomotive squeezing for quantum optical noises and quantum squashing for classical laser noise,  in general when a single quadrature is measured in homodyne detection.  Although interesting, these phenomena are not the focus of the present work.\\

\subsubsection{Tests of the analytical expression} 
As a check of the  the expression obtained above  we compare cavity heterodyne output PSDs obtained from Eq.(\ref{Shet}) with those obtained numerically.  The results, 
for a  $N=4$ particle, $N=4$ tweezer system,   are presented  in Fig.\ref{Fig4}.   

The pressure was taken to be $10^{-6}$ mbar and the temperature $T=300$ K.    The unperturbed frequencies were chosen as near-degenerate pairs to be $\omega_j/(2\pi)= 300. , 301.5, 285. , 286.5$ kHz,  while the coupling rates  were chosen to be close, but non-identical $g_j/(2\pi) 40, 38, 36, 34$ kHz.
We take $\Delta/(2\pi) =450$ kHz, while $\kappa/(2\pi) =390$ kHz.

The calculations were performed first for all tweezer phases equal; and then with tweezer phases all different from each other.  In the former case,  the  displacement mode $\hat{Q}(\omega)$  that couples to the light is real and corresponds to the
centre of mass mode (weighted by the couplings).  In the second case, the $\hat{Q}(\omega)$ is non Hermitian.
In both cases, the analytical expression is in perfect agreement (to within numerical precision)  with the numerics.

However, the spectra for the case of different phases differ substantially from the equal phase case.  In particular,  there is a significant shift of the CBM,  relative to the MMC. This indicates that the phases may offer a useful additional means of control. \\

Fig.\ref{Fig5} shows that many of the features analysed here  are already accessible with current 2-particle , 2D levitated optomechanics (2 particles in a cavity , with similar coupling in their $x,y$ modes),  thus a system with 4 displacement modes.   A two-tweezer, two particle cavity levitation experiment was recently reported \cite{Novotny2pcle}.
The figure illustrates the CBM, the MMC,  as well as the tendency (see right panels) of MMC modes to fill spectral gaps.

\section{Force sensing: signal to error considerations} 

It is interesting to consider the contribution to the measured displacement from an additional deterministic external force that acts on the levitated nanoparticles and represents an additional contribution in Eq. \ref{Qfnoises}:

\begin{align}
\label{Qfnoises2}
    \hat{Q}(\omega)
    \to
    \hat{Q}(\omega)
    +
    \sum_j
    \chi_j^\text{eff}(\omega)
    F_j^\text{ext}(\omega)
\end{align}

\begin{align}
    S_{QQ}^\text{\text{meas}}(\omega)
    \approx
    S_\text{BA}(\omega)
    +
    S_\text{\text{imp}}(\omega)
    +
    \sum_j
    S_{QQ}^{(j, \text{ext})}(\omega)
\end{align}

For a force $S_{FF}(\omega)$ that acts equally on all nanospheres, such as gravity, we can write:
\begin{align}
    \sum_j S_{QQ}^{(j, \text{ext})}(\omega)
    & \approx
    S_{FF}(\omega)
    \left|
        \sum_j \chi_j^\text{eff}(\omega)
    \right|^2
    \\
    & \approx
    S_{QQ}^\text{\text{meas}}(\omega)
    -
    (S_\text{BA}(\omega) + S_\text{\text{imp}}(\omega))
\end{align}
so
\begin{align}
    S_{FF}(\omega)
    \approx
    \frac{S_{QQ}^\text{\text{meas}}(\omega)}{|\chi_\text{eff}(\omega)|^2}
    -
    \frac{S_\text{BA}(\omega) + S_\text{\text{imp}}(\omega)}{|\chi_\text{eff}(\omega)|^2}
\end{align}
with
$\chi_\text{eff}(\omega) = \sum_j \frac{\mathcal{F}_j(\omega)}{\mathcal{M}_N(\omega) } \chi_j(\omega)$.

For the case where the particles are identical
$\Omega_j \equiv \omega_1$, $\Phi_j=0$, $g_j \equiv g$,
then

\begin{align}
    \chi_\text{tot}(\omega)
    \equiv
    N
    \frac{\chi_1(\omega)}{\mathcal{M}_N(\omega)}
\end{align}
and although nominally there is an enhancement factor of $N$ to the usual susceptibility, the denominator $\mathcal{M}_N$
is not advantageous, and accounts for strong damping of the mode, which for the `identical particles, identical modes' case turns a sharp resonance into the broad CBM.

A key issue in force sensing in optomechanics is that the susceptibility suppresses imprecision contributions near resonance,
where it is a maximum.  We consider the optimal case where the thermal gas component is negligible (high vacuum) thus $\Gamma_j \to 0$. 

In an actual experiment, heterodyne detection involves interference with a reference oscillator of frequency $\Omega_\text{het}$ and the actual output spectrum is
$S_{a_\text{out} a_\text{out}}(\Omega_\text{het} - \omega) + S_{a^\dagger_\text{out} a^\dagger_\text{out}} (\omega + \Omega_\text{het})$,
but this minor modification and frequency shift is straightforward  and not required for the present discussion  of force sensing.   The prefactor elucidates the dependence on $P_L$, the laser power, since $\bar{g}^2 \propto P_L$.  Increased power increases the back-action error but mitigates imprecision errors.

\section{Discussion and conclusions}
 We have shown that the spectra of a multiparticle array in a cavity has a characteristic  structure,   comprising a 
 broad strongly damped CBM (collective bright mode) ,  but also a `comb' of narrower frequency peaks,  the MMC. (multimode mechanical comb).  While the CBM has been identified in previous studies that considered $N$ identical nanoparticles,  with equal couplings and phases,  the MMC is a new generic feature identified and investigated here.  It is a phenomenon found in realistic arrays, where there are slight differences in the natural frequencies and couplings of the $N$ nanoparticles.
 
 We have shown that although a system of multiparticles trapped in different tweezers in a cavity can have a very large number of motional modes,  and the MMC has an intricate structure,  analysis of the cavity output is actually very simple. 
 It reduces to an effective 1D problem,  where the spectral PSDs are readily written down analytically,  and evaluated in closed form {\em for arbitrary $N$ and number of modes},  provided we consider an effective non-Hermitian coordinate.  
 
 In fact we show that the PSDs for the cavity output and displacements take a very similar form to the 
 well studied case of a single mechanical mode coupled to a single electromagnetic mode. The only difference is that the usual terms   are scaled by  functions $|\mathcal{F}_j(\omega)|^2, |\mathcal{F}_a(\omega)|^2 ,|\mathcal{F}_{a^\dagger}(\omega)|^2 $ and a modified normalization,  that are very simple to evaluate yet,  remarkably,  describe all the complexity of the multimode, multi-particle array spectrum.

 For simulating force sensing,  we need not concern ourselves with the complex displacement coordinate $\hat{Q}$, 
 and may simply relate the optical output to the external force:
  
  \begin{equation}
  \frac{S_{ \hat{a}_{out} \hat{a}_{out}} (\omega)}{ \kappa {g}^2 |\chi_c(\omega)|^2} \sim    | \chi^{\text{eff}}(\omega)|^2  S^{\text{ext}}_{FF}
  \end{equation}
where 
\begin{equation}
\chi_\text{eff}(\omega) = \sum_j \frac{\mathcal{F}_j(\omega)}{\mathcal{M}_N(\omega) } \chi_j(\omega)
\end{equation}

In conclusion,  the above analysis will simplify analysis of many experimental results involving arbitrarily large 
numbers of particles that all couple to a single cavity mode.  If the individual tweezer phases can be controlled individually 
this will also enable optimisation of the force susceptibilities and detection of signals over a broad band of frequencies.

{\em Acknowledgements:} Julian Iacoponi acknowledges an Engineering and Physical Sciences Research Council (EPSRC) DTP studentship. The authors acknowledge insightful discussions with Jayadev Vijayan and Peter Barker.  Funding is acknowledged from EPSRC Grant No. EP/W029626/1 for TM, 
and  an EPSRC-funded Quantum technology fellowship (grant number EP/S021582/1)  for MR.
\bibliographystyle{unsrt}
\bibliography{2Dbiblio}

\clearpage
\onecolumngrid

\setcounter{equation}{0}
\setcounter{figure}{0}
\setcounter{table}{0}
\setcounter{section}{0}
\setcounter{page}{1}
\makeatletter
\renewcommand{\thesection}{S\arabic{section}}
\renewcommand{\theequation}{S\arabic{equation}}
\renewcommand{\thefigure}{S\arabic{figure}}
\renewcommand{\bibnumfmt}[1]{[S#1]}
\renewcommand{\citenumfont}[1]{S#1}

\end{document}